\documentclass[12pt]{article}

\makeatletter
\def\appendix{\par\clearpage
  \setcounter{section}{0}
  \setcounter{subsection}{0}
  \@addtoreset{equation}{section}
  \def\@sectname{Appendix~}
  \def\theequation{\thesection.\arabic{equation}}
  \def\thesection{\Alph{section}}}
\makeatother

\usepackage{amssymb}
\usepackage{graphicx}

\begin{document}

\begin{titlepage}

\phantom{.}
\vspace{-2cm}

\hfill \vbox{\hbox{DFCAL-TH 03/2}
\hbox{September 2003}}

\vskip 0.8cm

\centerline{\bf Ultra-high energy neutrino-nucleon interactions}

\vskip 0.3cm

\centerline{R.~Fiore$^{a\dagger}$, L.L.~Jenkovszky$^{b\ddagger}$,
A.~Kotikov$^{c\star}$, F.~Paccanoni$^{d\ast}$, A.~Papa$^{a\dagger}$,
E.~Predazzi$^{e\$}$}

\vskip 0.3cm

\centerline{$^{a}$ \sl  Dipartimento di Fisica, Universit\`a della Calabria}
\centerline{\sl Istituto Nazionale di Fisica Nucleare, Gruppo collegato di Cosenza}
\centerline{\sl I-87036 Arcavacata di Rende, Cosenza, Italy}

\centerline{$^{b}$ \sl  Bogolyubov Institute for Theoretical Physics}
\centerline{\sl Academy of Sciences of Ukraine}
\centerline{\sl UA-03143 Kiev, Ukraine}

\centerline{$^{c}$ \sl Bogolyubov Laboratory of Theoretical Physics}
\centerline{\sl Joint Institute for Nuclear Research}
\centerline{\sl RU-141980 Dubna, Russia}

\centerline{$^{d}$ \sl Dipartimento di Fisica, Universit\`a di Padova}
\centerline{\sl Istituto Nazionale di Fisica Nucleare, Sezione di Padova}
\centerline{\sl via F. Marzolo 8, I-35131 Padova, Italy}

\centerline{$^{e}$ \sl Dipartimento di Fisica Teorica, Universit\`a di Torino}
\centerline{\sl Istituto Nazionale di Fisica Nucleare, Sezione di Torino}
\centerline{\sl via P. Giuria 1, I-10125 Torino, Italy}

\vskip 0.1cm

\begin{abstract}
Ultra-high energy ($E_\nu>10^8 $ GeV) neutrino-nucleon total cross
sections $\sigma^{\nu N}$ are estimated from a soft non-perturbative 
model complemented by an approximate QCD evolution, explicitly calculated. 
The resulting asymptotic energy behavior of the neutrino-nucleon 
cha\-rged-current total cross section is derived analytically and 
predictions concerning the observation of ultra-high energy neutrinos 
in future experiments are presented.
\end{abstract}

\vskip 0.1cm
\hrule

\vfill

$
\begin{array}{ll}
^{\dagger}\mbox{{\it e-mail address:}} &  \mbox{fiore,~papa@cs.infn.it} \\
^{\ddagger}\mbox{{\it e-mail address:}} & \mbox{jenk@gluk.org} \\
^{\star}\mbox{{\it e-mail address:}} & \mbox{kotikov@thsun1.jinr.ru} \\
^{\ast}\mbox{{\it e-mail address:}} & \mbox{paccanoni@pd.infn.it} \\
^{\$}\mbox{{\it e-mail address:}} & \mbox{predazzi@to.infn.it}
\end{array}
$

\end{titlepage}
\eject
\newpage

\section {Introduction} 

Calculations of ultra-high energy (UHE) neutrino-nucleon cross
sections $\sigma^{\nu N}$ \cite{YMA,KR,RG,KUS,KMS,MR,MG,RB} have 
attracted attention for many years since, on the one hand, UHE neutrino
fluxes are predicted by cosmological models but have not yet been
measured experimentally and, on the other hand, theoretical
models exist for the evaluation of these cross sections.

Neutrino telescopes are the most promising tools for probing the
distant stars and galaxies. Due to their high stability and
neutrality, neutrinos arrive at a detector on a direct line from
their sources, undeflected by intervening magnetic fields, with
expected energy up to $10^{20}$ eV. Whereas high-energy photons are
completely absorbed by a few hundred grams/cm$^2$ of material, the
interaction length of a 1 TeV neutrino is about 250
kilotonnes/cm$^2$~\cite{RG}.

An important scientific goal of large-scale neutrino telescopes is
the detection of UHE cosmic neutrinos produced outside the atmosphere, 
i.e. neutrinos produced by galactic cosmic rays interacting with 
interstellar gas, and extra-galactic neutrinos.
High neutrino energy brings a number of advantages~\cite{RG,KUS}.
Firstly, the charged-current cross section increases as
$\sigma\sim E_{\nu}$ for $E_{\nu}<10^{12}$ eV and is believed to
rise as $\sigma\sim E_{\nu}^{0.4}$ for higher $E_{\nu}$. (In
this paper we question the last point.) Secondly, the background
of atmospheric neutrinos decreases as compared to the flux from
extra-galactic sources, approximately as $E^{-1.6}_{\nu}.$

Another important and promising aspect of UHE neutrino physics is
the possibility to study the production of exotic objects, such as
supersymmetric particles~\cite{BKK} or black holes~\cite{RG}.

The phenomenology of UHE neutrinos and their detection depend on
the neutrino-nucleon total cross section, which has been calculated
in the standard model (see e.g.~\cite{RG}). A striking feature of
these calculations is the continued power-law growth of the
cross sections with $E_{\nu}$ at highest energies. The rise with
$E_{\nu}$, on the other hand, is directly related to the very
low-$x$ behavior of the nucleon parton distributions derived 
from e.g. the HERA data. Problems related to the unitarity bound 
have been discussed in~\cite{MR,DAD,KK03,J-M03}.

At ultra-high neutrino energies the total cross section is
completely dominated by its deep inelastic component. The energy
dependence of neutrino-nucleon cross section, as suggested from
the small-$x$ nucleon structure functions (SF) measured at HERA, should
show a rapid increase with energy, like $s^{\lambda}$~\cite{power}, 
where $\lambda$ scales as $\ln(Q^2/\Lambda^2)$ and ranges approximately 
in the interval $0.1\div 0.4$ for $1$ GeV$^2 < Q^2 < 100$ GeV$^2$ (see
Ref.~\cite{H1}).

The highest available $ep$ energy at HERA, $\sqrt{s_{ep}}=314$
GeV, is still too low as compared to the expected UHE neutrino-nucleon
collision energies. For the neutrino energy $E_\nu\sim10^{12}$
GeV, the relevant $x$-region is $x\sim 10^{-8}$, 
while the HERA measurements extend now to $x\sim 10^{-6}$; however such low 
values of $x$ are measured at $Q^2<0.1$ GeV$^2$. For the interesting region 
of $Q^2\sim M^2_W$, SF are measured only up to $x\sim10^{-3}$ in the D0 
experiment of inclusive jets~\cite{BAB}. 
Since studies of the UHE $\nu N$ cross sections began nearly a
decade ago~\cite{YMA,KR,BKK,FKR} our knowledge about quark
distributions and nucleon SF improved significantly~\cite{fits}.

In the present paper we study the UHE behavior of the neutrino-nucleon 
charged-current cross section starting from models where nucleon SF have a 
milder increase. 
Our analysis relies strongly on recent progress and our previous experience 
in the studies of the nucleon SF~\cite{fits,ELBA,DGJ,Rivista,DJLP,CAL,JLP}, 
based on phenomenological fits, extended by QCD evolution.

The paper is organized as follows: in the next Section we briefly review the
steps involved in the determination of the neutrino-nucleon cross section
and illustrate our approach; in Section~3 we study the asymptotic behavior 
of the total neutrino-nucleon cross section; in Section~4 we 
discuss the implications of our results for the observation of UHE neutrinos
in future experiments and finally in Section~5 we draw some conclusions.

\section {Nucleon structure functions and neutrino-nucleon cross sections}

In order to compute the cross section for neutrino-nucleon
interactions at high energies, one needs a detailed description of
the quark structure of the nucleon. The special challenge of the
UHE neutrino-nucleon cross section is that the $W$-boson
propagator emphasizes smaller and smaller values of $x$ as the
neutrino energy $E_{\nu}$ increases. In the UHE domain, the most
important contribution to the $\nu N$ cross section comes from
$x\sim M^2_W/(2ME_{\nu})$ and hence the greatest uncertainty in the
predictions comes from the small-$x$ extrapolation.

Deep inelastic scattering of UHE neutrinos is unique in the sense
that it explores extreme regions of the $(Q^2,x)$ phase space
where no accelerators data exist. One thus must rely on
predictions or assumptions regarding the nature of these
distributions, in particular in the region of ultra-low $x$ and
low $Q^2$, as well as in the region of low $x$ and low $Q^2.$

Let us consider the double differential cross section for the deep
inelastic scattering of neutrinos on an isoscalar target for 
charged-current processes. We will limit ourselves
to the leading order corrections to the simple parton model and
hence all parton model formulas remain unchanged except that the
parton distributions now depend on $x$ and $Q^2$ and not only on
$x$. In particular, the Callan-Gross relation, $F_2=2xF_1$ or
$F_L=0$, holds in leading order~\cite{BU}. With the usual notation
we can then write for charged-current neutrino interactions
\begin{eqnarray}
\left(\frac{d\sigma}{dx\,dy}\right)^{\nu} & = &
\frac{G_F^2ME_{\nu}}{\pi} \left(
\frac{M_W^2}{Q^2+M_W^2}\right)^2 \nonumber \\
&\times& \left\{\frac{1+(1-y)^2}{2}F_2^{\nu}(x,Q^2)+
\left(1-\frac{y}{2}\right)yxF_3^{\nu}(x,Q^2)\right\}. \label{n1}
\end{eqnarray}
For anti-neutrino charged-current processes one must change the
sign in front of $F_3^{\nu} (x,Q^2)$. (The changes necessary in
order to describe neutral-current neutrino interactions are given
in many textbooks and review articles, for example in~\cite{BU}).
In Eq.~(\ref{n1}), $G_F$ is the Fermi constant, $M$ is the nucleon
mass and the variable $y$ is related to the Bjorken $x$ through
the relation
\begin{equation}
y=\frac{Q^2}{x(s-M^2)}\simeq \frac{Q^2}{xs}\;, \label{n2}
\end{equation}
where $s$ is the square of the c.m. energy for the
neutrino-nucleon  scattering. The laboratory neutrino energy
$E_{\nu}=(s-M^2)/(2M)$ is approximately $s/(2M)$ in the energy region
of interest. 

The main purpose of this paper is to evaluate the asymptotic behavior 
of the total cross section for charged-current neutrino-nucleon process
\begin{equation}
\sigma^{\nu N\,(CC)}=\int_0^1
dx\,\int_0^1dy\,\left(\frac{d\sigma}{dx\,dy}\right)^{\nu}
\label{n3}
\end{equation}
in a model for the structure function $F_2$ based on a soft
non-perturbative input complemented by an approximate QCD
evolution. The contribution of $xF_3$ can in fact be neglected at
ultra-high energies. Unfortunately, experimental data for the
neutrino SF do not extend to the ``deep sea'' region, i.e. to 
$x$ smaller than 0.001, and to large $Q^2$. Both these limits, 
for the electromagnetic structure functions $F^{ep}$, are subject of 
debate regarding the presence of possible saturation effects at 
small $x$ and discrepancies in the $Q^2$ evolution for large $Q^2$. 
$F_2^{\nu}$ differs from $F_2^{ep}$ at small $Q^2$ as can be 
easily seen by formally integrating the double differential cross 
section (\ref{n1}) over $x$ at fixed value of the product $(E_{\nu} y)$
\begin{equation}
\lim_{y\to
0}\frac{1}{E_{\nu}}\frac{d\sigma}{dy}=\frac{G_F^2M}{\pi}
\int_0^1dx\,F_2^{\nu}(x,Q^2=0)\;, \label{n4}
\end{equation}
formula that has been used for the normalization of the neutrino
flux~\cite{CCFR1}.  The integral in Eq.~(\ref{n4}), when evaluated for
$F_2^{ep}$, would vanish by gauge invariance. At increasing $Q^2$,
this difference will disappear but, at the moment, no smooth
transition has been found in all the range of the variables
appearing in Eq.~(\ref{n3}).

\subsection{The non-perturbative input}

In a number of papers~\cite{ELBA,DGJ,Rivista,DJLP,CAL,JLP,COSENZA} 
a diffraction model has been developed, describing equally well
hadronic reactions at high energies and DIS at small $x$. The main
idea behind the model is that SF at small and moderate values of
$Q^2$ are Regge-behaved. The $Q^2$-dependence was introduced in a
phenomenological way.

It was shown~\cite{ELBA,DGJ,DJLP,JLP,COSENZA} that data on deep
inelastic $ep$ scattering from HERA and elsewhere can be fitted
perfectly well by a ``soft'' Pomeron alone. This Pomeron could be
a dipole with a $\ln(1/x)$ dependence or a squared logarithm,
saturating asymptotically the Froissart bound. For the
electromagnetic SF, parametrizations exist~\cite{fits} that allow 
for a simplified evolution until very large $Q^2$, while maintaining 
a soft dependence on the Bjorken $x$. As an example we will consider 
a structure function $F_2^{\nu}(x,Q^2)$ of the form
\begin{equation}
F_2^{\nu}(x,Q^2)= \left[
a(Q^2)+b(Q^2)\ln\left(\frac{1}{x}\right)\right]
(1-x)^{\nu(Q^2)} \label{n5}
\end{equation}
and derive for it an explicit approximate evolution starting from
$Q^2=Q_0^2$. The main motivation for our choice 
of the model is that at small $Q^2$ there is similarity between the small-$x$ 
behavior of the structure function and the $s$-dependence, $x\sim Q^2/s$, 
of the hadronic total cross section at high energies. The alternative approach, 
with a power behaved dependence, has been discussed in the literature in several
publications. The differences between the two approaches can be
significant and will be discussed in Section~3.

We notice that the coefficients $a(0)$, $b(0)$ and $\nu(0)$ are constrained by
Eq.~(\ref{n4}) whose left hand side can be determined
from the experiment, Ref.~\cite{CCFR1}. The constraint is
\begin{displaymath}
\lim_{y\to 0}\frac{1}{E_{\nu}}\frac{d\sigma}{dy}=
\frac{1}{\nu(0)+1}\biggl\{a(0)+b(0)\biggl(
\gamma+\psi[\nu(0)+2]\biggr)\biggr\}\;,
\end{displaymath}
where $\gamma$ denotes the Euler's constant and $\psi[z]$ is the
logarithmic derivative of the $\Gamma$-function. When low-$Q^2$
data~\cite{FLEMING} will extend to lower $x$ values, it will be
possible to check the choice~(\ref{n5}). 

By imposing a lower cut in $Q^2$, $Q^2=Q_0^2$, and rewriting Eq.~(\ref{n3}) as
\begin{displaymath}
\sigma^{\nu N\,(CC)}\simeq\frac{1}{2ME_{\nu}}\int_{Q_0^2}^s\,dQ^2\,
\int_{Q^2/s}^{1}\,\frac{dx}{x}\left(\frac{d\sigma}{dx\,dy}\right)^{\nu}\;,
\end{displaymath}
where in the differential cross section $y=Q^2/(xs)$, the $F_2$
contribution to  the total cross section can be written in the
form
\begin{displaymath}
\bar{\sigma}^{\nu N}\equiv\frac{\sigma^{\nu
N}+\sigma^{\bar{\nu}N}}{2}
\end{displaymath}
\begin{equation}
=\frac{G_F^2}{2\pi}\int_{Q_0^2}^s\,dQ^2\left(
\frac{M_W^2}{Q^2+M_W^2}\right)^2
\int_{Q^2/s}^{1}\,\frac{dx}{x}\;\frac{1+(1-Q^2/(xs))^2}{2}F_2^{\nu}(x,Q^2)\;.
\label{n6}
\end{equation}

\subsection{$Q^2$ evolution}

Let us consider a simplified approach to the DGLAP~\cite{DGLAP}
evolution, where the parton distribution functions satisfy the
relation $q(x)=\bar{q}(x)=u(x)=d(x)=s(x)=2 c(x)=2b(x)$ and
$F_2^{\nu}$, for five flavors, is given by
$F_2^{\nu}=8xq(x)$. This assumption gives $F^{ep}_2=17 x q(x)/9$ and, 
as shown in~\cite{FKR}, with this simple form for the SF a high quality fit to 
the HERA data can be obtained.

For each parton distribution we assume the
same function (\ref{n5}) chosen for $F_2^{\nu}(x,Q^2)$
\begin{equation}
xf_i(x,Q^2)=\left[a_i(Q^2)+b_i(Q^2) \ln\left(\frac{1}{x}
\right)\right]\,(1-x)^{\nu(Q^2)},\;\;\;\;(i=q,g) \label{n7}
\end{equation}
with the same functional form for quark and gluon but with
different coefficients. Experimental data for quarks and global
analysis, for example the parton analysis of MRST~\cite{fits}, for
the gluon could determine the coefficients in the expressions for
$xf_i(x,Q^2)$, where $i=q,g$ stands for quark or gluon, respectively. 

In the leading order of perturbation theory, an approximate evolution is 
feasible by using the results of Ref.~\cite{KP}. The only
condition is that $a_i\gg b_i$ $(i=q,g)$, where $a_i\equiv
a_i(Q_0^2)$ and $b_i\equiv b_i(Q_0^2)$, so that no interference
appears in the $Q^2$ evolution of the coefficients multiplying the
different powers of the logarithm\footnote{The requirement $a_i\gg b_i$
is needed for all $Q^2$ values. It drastically simplifies our 
formulas since it avoids a mixing between $a_i$ and $b_i$ parameters 
during the $Q^2$ evolution.}. We define $\beta_0=11-2f/3$,
where $f=5$ is the number of flavors in our approach,
\begin{displaymath}
t=\ln\left[\frac{\alpha(Q_0^2)}{\alpha(Q^2)}\right]=\ln \left[
\frac{\ln(Q^2/\Lambda^2)}{\ln(Q_0^2/\Lambda^2)}\right]
\end{displaymath}
and recall the form of the scaling variables,
\begin{equation}
\sigma=2\sqrt{-\hat{d}_{gg}t\,\ln(1/x)},\;\;\;\;\;\;
\rho=\sqrt{\frac{-\hat{d}_{gg}t}{\ln(1/x)}}=\frac{\sigma}{2\ln(1/x)}\;,
\label{n8}
\end{equation}
where $\hat{d}_{gg}=-12/\beta_0$~\footnote{We follow here the
notation of~\cite{KP}: $d$ is the ratio between the anomalous
dimension and twice the QCD $\beta$-function in leading order,
while $\hat{d}$ and $\bar{d}$ are, respectively, the singular and
regular part of $d$ when $n\to 1$.} . Then, from~\cite{KP}, we get
for small $x$ the evolution of the structure function $F_2(x,Q^2)=
8xf_q(x,Q^2)=8x[f_q^-(x,Q^2)+f_q^+(x,Q^2)]$ as
\begin{eqnarray}
xf_q^-(x,Q^2) & = & \left(a_q+b_q\left[\ln\left(\frac{1}{x}\right)
-C_1(\nu)\right]\right)\cdot e^{-d_-(1)t}\;+O(x) \label{n9} \\
xf_q^+(x,Q^2) & = & \left(a_q^+\rho I_1(\sigma)+b_q^+I_0(\sigma)
\right)\cdot e^{-\bar{d}_+(1)t}\;+O(\rho), \label{n10}
\end{eqnarray}
where $d_-(1)=16 f/(27\beta_0),\,\bar{d}_+(1)=1+20 f/(27\beta_0)$,
$a_q^+=f(a_g+4 a_q/9)/9$ and an analogous expression for $b_q^+$.
$C_1(\nu)$ is related to the logarithmic derivative of the
$\Gamma$-function: $C_1(\nu)=\psi[1+\nu]-\psi[1]$. $I_0(z)$ and $I_1(z)$ 
are modified Bessel functions of the first kind.

We notice that the presence of higher powers of the logarithm of $1/x$ can
be easily handled in this formalism provided that $a_i\gg b_i \gg\ldots$. 
While the evolution given by Eqs.~(\ref{n9})-(\ref{n10})
is correct in the limit of very large $\sigma$ and $Q^2
\gg Q_0^2$, the requirement that the initial input must be
reproduced for $Q^2=Q_0^2$ introduces a corrective term in
$f_q^+$ that does not affect the evolution for large $Q^2$.
Eq.~(\ref{n10}) should then be replaced by
\begin{equation}
xf_q^+(x,Q^2)=\left(a_q^+\rho
I_1(\sigma)+b_q^+\left[I_0(\sigma)+\Delta\right] \right)\cdot
e^{-\bar{d}_+(1)t}\;+O(\rho)\;, \label{n11}
\end{equation}
where $\Delta=b_qC_1(\nu)/b_q^+-1$.

\section{Asymptotic behavior of $\sigma^{\nu N}$}

We rewrite Eq.~(\ref{n6}) in the form
\begin{displaymath}
\bar{\sigma}^{\nu N}=\frac{2G_F^2}{\pi}\int_{Q_0^2}^s\,dQ^2\left(
\frac{M_W^2}{Q^2+M_W^2}\right)^2 
\end{displaymath}
\begin{equation}
\times\int_{Q^2/s}^{1}\,\frac{dx}{x}\;\left[1+
\left(1-\frac{Q^2}{xs}\right)^2\right]
\left\{xf_q^-(x,Q^2)+xf_q^+(x,Q^2)\right\}\;, \label{n12}
\end{equation}
where $xf_q^{\mp}(x,Q^2)$ are given in Eqs.~(\ref{n9})-(\ref{n10})
respectively. The integral over $x$ can be done exactly for
$f_q^-(x,Q^2)$. Let us denote this integral with the symbol
$J_-(s,Q^2)$. The result is presented in Eq.~(\ref{a1}) of
the Appendix, and the leading behavior, for large $s$, of the
$x$-integral is given by $\ln^2(s/Q^2)$. 

The presence of many scales $(\Lambda^2,\,Q_0^2,\,M_W^2\,\mbox{and}\;s)$ 
makes difficult any precise estimate of the asymptotic behavior of the
$xf_q^-(x,Q^2)$ contribution to $\bar{\sigma}^{\nu N}$. From the
Ostrowski's inequality~\cite{GR} for integrals it follows, however, that
this contribution has an upper bound for $s \to\infty$:
\begin{equation}
\left|\int_{Q_0^2}^s\,dQ^2\left( \frac{M_W^2}{Q^2+M_W^2}\right)^2
J_-(s,Q^2)\right| \leq\;K\ln^2\left(\frac{s}{Q_0^2}\right)\;,
\label{n13}
\end{equation}
where $K$ is a suitable constant. 

Only the small-$x$ evolution of $f_q^+(x,Q^2)$ is known from (\ref{n10}) 
and the condition $\rho \ll 1$ requires a cutoff at the upper limit of the
$x$-integral. Let us call it $x_M$ with $x_M < 1$, while the
$Q^2$-integral has now $x_Ms$ as the upper limit of integration.
The first integral we meet, when considering the term with
$f_q^+(x,Q^2)$, is
\begin{equation}
\int_{Q^2/s}^{x_M}\frac{dx}{x}\;xf_q^+(x,Q^2)=e^{-\bar{d}_+(1)t}
\left(a_q^+{\cal I}_1+b_q^+{\cal I}_2\right) \label{n14}
\end{equation}
and, with the change of variable $x=\exp[-z^2/(-4\hat{d}_{gg}t)]$,
the solution can be easily found. Setting $\sigma_u=2\sqrt{-\hat{d}_{gg}t
\ln(s/Q^2)}$ and $\sigma_{\ell}=2\sqrt{-\hat{d}_{gg}t\ln(1/x_M)}$, we obtain:
\begin{equation}
{\cal I}_1(s,Q^2)=\int_{\sigma_\ell}^{\sigma_u}\,dz\,I_1(z)
=I_0(\sigma_u)-I_0(\sigma_\ell)
\label{n15}
\end{equation}
and
\begin{equation}
{\cal I}_2(s,Q^2)=
\frac{1}{-2\hat{d}_{gg}t}\int_{\sigma_\ell}^{\sigma_u}\,dz\,zI_0(z)
=\frac{1}{-2\hat{d}_{gg} t}\biggl(\sigma_u \,I_1(\sigma_u)-
\sigma_\ell \,I_1(\sigma_\ell)\biggr)\;.
\label{n16}
\end{equation}
Other integrals, coming from powers of $Q^2/(xs)$ in Eq.~(\ref{n12}), 
are $O(\rho)$ with respect to those in Eqs.~(\ref{n15}) and (\ref{n16}) 
and it is consistent with our approximation in Eq.~(\ref{n10}) to neglect 
them. The proof can be found in the Appendix. 

We define now the integrals
\begin{equation}
{\cal J}_k(s)=\int_{Q_0^2}^{x_Ms}\,dQ^2\left(
\frac{M_W^2}{Q^2+M_W^2} \right)^2 e^{-\bar{d}_+(1)t}{\cal
I}_k(s,Q^2),\;\;\;\;(k=1,2) \label{n17}
\end{equation}
and notice that their large-$s$ behavior determines completely the
$f_q^+$ contribution to the mean total cross section in this
limit. Of the two integrals ${\cal J}_1(s),\,{\cal J}_2(s)$ only
the first one needs to be evaluated explicitly since
\begin{displaymath}
\frac{d{\cal J}_2(s)}{d\ln s}={\cal J}_1(s)+O\left( \frac{1}{\ln
s}\right)\;.
\end{displaymath}
We consider then the integral ${\cal J}_1(s)$. The function
under the integral sign in ${\cal J}_1(s)$ vanishes at both
integration limits and has a bell-shaped form with a well defined
maximum. The maximum moves at larger values of $Q^2$, when $s$
increases, and its position is defined, in the limit $s\to
\infty$, by the solution of the transcendental equation
\begin{displaymath}
\beta_0 \alpha_s(Q_0^2)\ln(s/Q^2)- \sqrt{\frac{t
\ln(s/Q^2)}{-\hat{d}_{gg}}}
\end{displaymath}
\begin{equation}
\times\left(\frac{8\pi Q^2}{M_W^2+Q^2}e^{t}+ \beta_0\alpha_s(Q_0^2)
\bar{d}_+(1)\right)-4 \pi e^t=0 \;,
\label{n18}
\end{equation}
where the variable $t$ has been rewritten as
\begin{displaymath}
t=\ln\left[1+ \frac{\beta_0 \alpha_s(Q_0^2)}{4\pi}\,\ln(Q^2/Q_0^2)\right]\;.
\end{displaymath}
Eq.~(\ref{n18}) can be solved for $\ln(s/Q^2)$, obtaining an
explicit expression for $s$ as a complicated function of the value
of $Q^2$ at the maximum $Q^2_{\mbox{\scriptsize max}}$. If we rewrite 
Eq.~(\ref{n18}) as $f(s,Q^2)=0$, we get the slope
\begin{displaymath}
\frac{d Q^2_{\mbox{\scriptsize max}}}{ds}=-\left.\frac{\frac{d f(s,Q^2)}{d
s}}{\frac{d f(s,Q^2)}{d Q^2}} \right|_{Q^2=Q^2_{\mbox{\scriptsize max}}}
\end{displaymath}
that, as can be easily shown, vanishes when $s \to\infty$.

In order to estimate the position of the maximum as a
function of $s$, in the range $10^7$ GeV $\leq s \leq 10^{15}$ GeV, we
have chosen the parameters in intervals that include their most
probable value and found that the position of the maximum can be
roughly reproduced by a term linear in $(\ln s)^{1.5}$, while the
value of the integrand, at the maximum, is near the value of
the Bessel function there. The width at half height of the peak
goes like $\ln s$. By multiplying the value, at the maximum,
of the integrand in ${\cal J}_1(s)$ by the width of the peak, 
we get the right order of magnitude for the corresponding
integral in Eq.~(\ref{n17}), as can be verified by a numerical
computation. In this limit the mass of the $W$-boson plays little
role since $Q^2_{\mbox{\scriptsize max}}\ll  M_W^2$ until $s\approx 10^{15}$
GeV$^2$. It is then possible to conclude that, in this approach,
the total cross section for neutrino-nucleon scattering increases
with $s$ faster than any power of $\ln s$ but slower than any power of
$s$. 

\begin{figure}[tb]
\centering
\includegraphics[width=\textwidth]{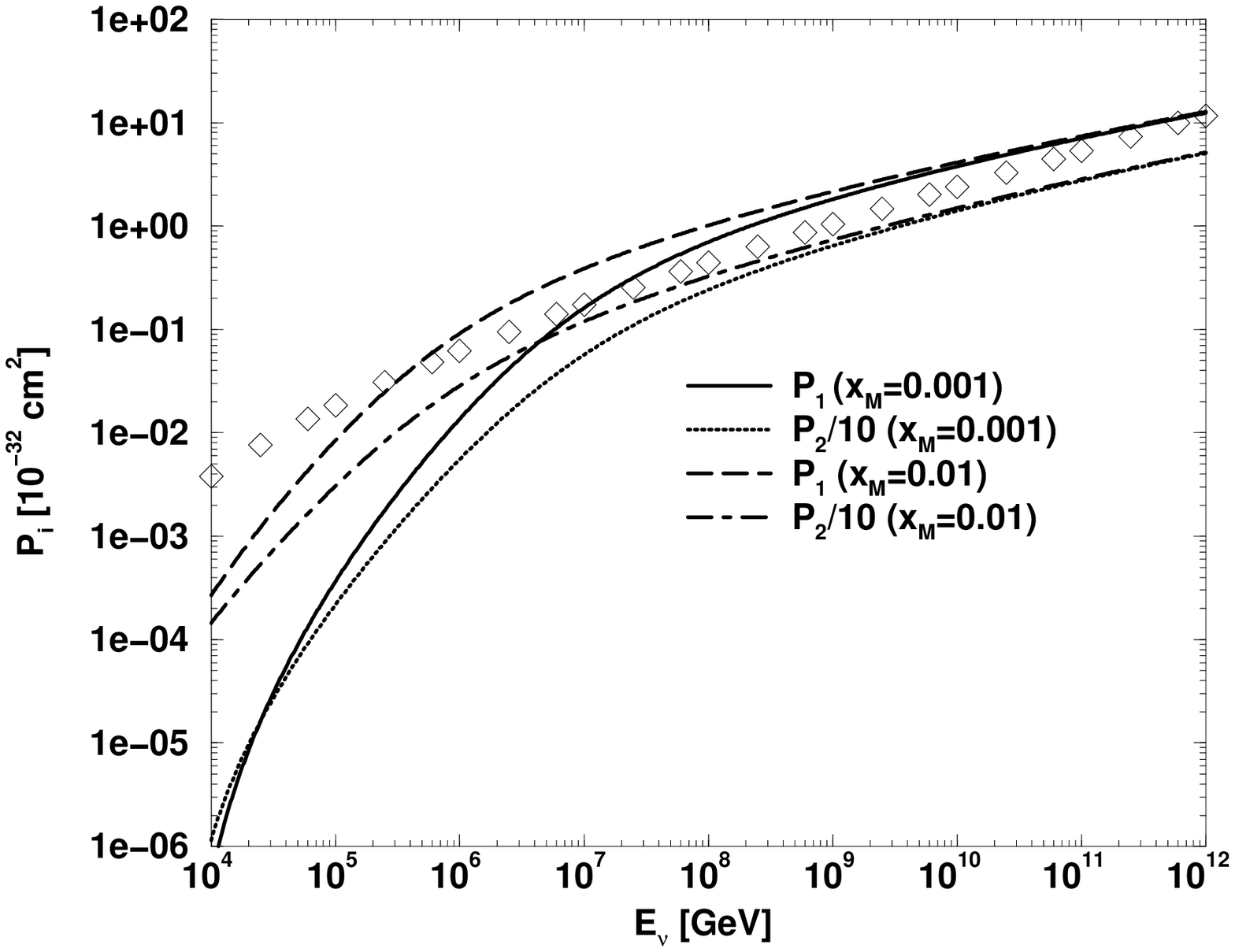}
\caption[]{\small The curves of $P_1$ (solid for $x_M=0.001$, dashed
$x_M=0.01$) and $P_2/10$ (dotted for $x_M=0.001$, dot-dashed for
$x_M=0.01$) are plotted versus the neutrino energy $E_\nu$ together
with data for $\bar{\sigma}^{\nu N}$ (diamonds) from Ref.~\cite{RG}.}
\label{fig1}
\end{figure}

Most of the existing models of $F_2$ assume a fast, power-like increase 
in $1/x$ that transforms in a like increase of 
$\sigma^{\nu N}(E)$~\cite{YMA,KR,RG,KUS,KMS,MR,MG,FKR,Gaz}. In an alternative 
approach~\cite{ELBA,DGJ,Rivista,DJLP,CAL,JLP,RW} 
the HERA data are shown to fit a model of the SF, at low and
intermediate $Q^2$ with a moderate, logarithmic, increase in
$1/x$, the observed ``HERA effect'' being attributed to the decrease
of the non-leading contributions, relevant at larger $x$. DGLAP
evolution transforms the logarithmic behavior in the way we have
shown in Eqs.~(\ref{n9})-(\ref{n10}). It is important, however, to 
realize that this asymptotic behavior is quite different from the 
one obtained starting with a
power, $x^{-\mu}$. According to~\cite{FKR,KP,AK}, if the starting
distributions have a power-law form in $x$, the leading term keeps
the same $x$-dependence under DGLAP evolution also at the
next-to-leading order. The component $xf_q^+(x,Q^2)$ in
Eq.~(\ref{n10}) can be represented, in the leading order of
perturbation theory and in the small-$x$ limit, by the
approximated form~\cite{AK,AK1}
\begin{equation}
xf_q^+(x,Q^2)\simeq \frac{f}{9}\mu\,xf_g(x,Q_0^2)e^{-d^+t}\;,
\label{n19}
\end{equation}
if $xf_i(x,Q_0^2)=h_i(Q_0^2)x^{-\mu},\;\;(i=q,g)$. In Eq.~(\ref{n19}),
$d^+=\gamma^+/(2\beta_0)$ is the largest eigenvalue of the
anomalous dimension matrix evaluated at $(1+\mu)$. \footnote{Note that the 
power-like behavior $\sim x^{-\hat d_{gg} t}$ of
parton distributions is in agreement also with the DGLAP dynamics 
in the $Q^2$ range near the starting point $Q^2_0$ (see Refs.~\cite{JKP,AK1}).
However, the power-like behavior cannot be combined together with~(\ref{n19}) 
to some unified Regge-like $Q^2$-evolution of parton distributions.}
For $\mu\sim 0.4$, as in Ref.~\cite{FKR}, $d^+$ is negative and
$d^+\approx -3.4$. In this approach it is easy to show that
$\bar{\sigma}^{\nu N}\propto (s/M_W^2)^{\mu}+O(s^0)$.

It is interesting to compare our result based on Eqs.~(\ref{n15}) and
(\ref{n16}) with a model where the cross section increases like a
power of $s$~\cite{RG}. We notice that Eq.~(\ref{n12}) can be written as
\begin{equation}
\bar{\sigma}^{\nu N}\approx \frac{4G_F^2}{\pi}\left(a_q^+{\cal
J}_1+b_q^+{\cal J}_2\right) \label{n20}
\end{equation}
and that all the integrals can be evaluated numerically, choosing for 
example $Q_0^2=10$ GeV$^2$, $\alpha_s(Q_0^2)=0.23$ and the values
$x_M=0.001,0.01$. The choice of $Q^2_0=10$ GeV$^2$ is reasonable because in the 
region of $Q^2$ near this value the rise in $x$ of SF is well
described by a $\ln(1/x)$ and/or $\ln^2(1/x)$ behavior
\footnote{For larger $Q^2$ values, the rise of $F_2$ is stronger and is
better described by a power law $x^{-\lambda}$, with
$\lambda \sim 0.3$ at $Q^2 \sim 100$ GeV$^2$. Evidently, our initial 
conditions~(\ref{n7}) cannot be applied here. However, after $Q^2$ evolution 
in our approach, we get a behavior with a $Q^2$-dependent effective slope 
$\lambda_{\mbox{\scriptsize eff}}(Q^2)$ in agreement with experimental 
data~\cite{H1}.}.

By evaluating the integrals numerically,
a comparison can be done with the values of $\bar{\sigma}^{\nu N}$
obtained from Table I of Ref.~\cite{RG}. We plot in 
Fig.~\ref{fig1} the values of $P_i\equiv (4G_F^2/\pi) {\cal J}_i$ 
and the result of Ref.~\cite{RG}.
As can be seen from Fig.~\ref{fig1}, the power-behaved cross section
increases with $s$ faster than the components of $\bar{\sigma}^{\nu
N}$ in the relation~(\ref{n20}). The explicit values of the parameters $a_q^+$
and $b_q^+$ do not enter this comparison, but it turns out that
their order of magnitude can be made consistent with the
parametrizations of the HERA data studied in Ref.~\cite{DJLP}.

The Bessel-like asymptotic behavior of the total cross section in 
Eq.~(\ref{n20}) violates the Froissart bound since it is steeper than any 
power of $\ln s$, though being flatter than any power of $s$. The 
comparison with a power-behaved cross section, strongly violating the 
Froissart bound, makes us confident on the fact that additional 
contributions needed to restore unitarity should be less relevant in 
our approach.

\section{Observation of ultra-hi\-gh energy neutrinos}

The evolution in leading order considered in the present paper overestimates 
the cross sections. Next-to-leading calculations show (see, for example, 
Ref.~\cite{RB}) that the ratio 
$\sigma_{\mbox{\scriptsize NLO}}/\sigma_{\mbox{\scriptsize LO}}$ decreases when
the neutrino energy increases and it is reasonable to assume that our estimates
constitute an upper bound for the true predictions of the model. Hence the NLO 
evolution of the soft non-perturbative input could lead to a more 
pronounced difference in the predictions with respect to a power-like 
input. Another possible cause for the decrease of the true cross sections,
that has not been taken into account here, resides in the gluon recombination
at small $x$. \footnote{The results of numerous analyses of the higher-twist 
contributions in DIS~\cite{GLR83} and of recent papers discussing screening 
effects in neutrino-nucleon cross-section~\cite{KK03,J-M03} are still not 
conclusive.}
This recombination can be mimicked by considering a large, 
``effective'' gluon anomalous dimension, with a consequent increase of the
value of $\bar d_+(1)$ that appears in the exponential $\exp(-\bar d_+(1)\,t)$
in Eq.~(\ref{n10}). Despite these possible corrections, it is of interest to 
consider the constraints imposed by our model on the interaction length of 
neutrinos and compare them with a more conventional approach. 

\begin{figure}[tb]
\centering
\includegraphics[width=\textwidth]{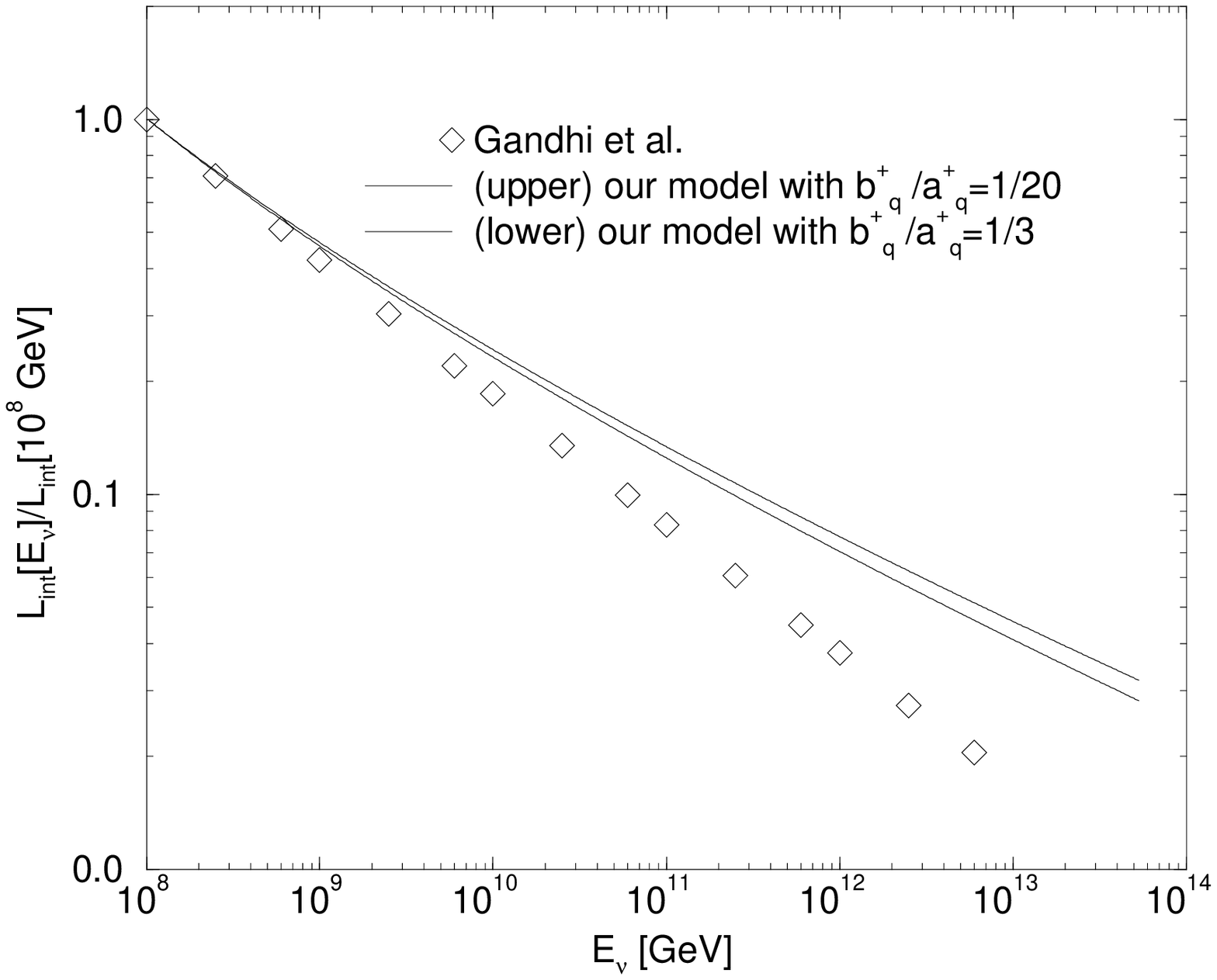}
\caption[]{\small 
$L_{\mbox{\scriptsize int}}(E_\nu)/L_{\mbox{\scriptsize int}}(10^8 \mbox{ GeV})$
versus $E_\nu$ in our model (solid lines) and according to 
the results of Ref.~\cite{RG} (diamonds).}
\label{fig2}
\end{figure}

For this purpose, we considered the (water equivalent) interaction length, 
defined as 
\[
L_{\mbox{\scriptsize int}}(E_\nu)=\frac{1}{\sigma^{\nu N}(E_\nu) N_A}\;,
\]
where $N_A=6.022\times 10^{23}$ cm$^{-3}$
(water equivalent) is the Avogadro's number, and 
evaluated the ratio of $L_{\mbox{\scriptsize int}}(E_\nu)$ to 
$L_{\mbox{\scriptsize int}}(10^8 \mbox{ GeV})$ as obtained from our model, 
with $b_q^+/a_q^+$ ranging between 1/20 and 1/3 and setting 
$Q_0^2=10$ GeV$^2$, $\alpha_s(Q_0^2)=0.23$ and $x_M=0.01$.
In Fig.~\ref{fig2} our result is compared with the same ratio determined
from the results of Ref.~\cite{RG}. At higher energies $L_{\mbox{\scriptsize int}}$
in our model could be larger than expected in more conventional
approaches.

Experiments are planned to detect UHE neutrinos by observation of
the nearly horizontal air showers (HAS) in Earth atmosphere
resulting from $\nu$-air interactions. The expected rates are
proportional to the neutrino-nucleon  cross section. The resulting
cross section at $10^{20}$ eV is rather high, $\sim 10^{-31}$ cm$^2$ 
according to the estimates of~\cite{FKR,RG}, and
results from the extrapolation of parton distribution functions
far beyond the reach of experimental data. 

If the cross section is lower, then the event rate for neutrino-induced
HAS is reduced by the same factor. This reduction would compromise
the main detection signal that has been proposed for UHE neutrino
experiments. A smaller cross section would however offer a double
advantage for the planned experiments~\cite{KUS}:
\begin{enumerate}
\item with a new search strategy, based on the observation of
upgoing air showers (UAS) initiated by muon and tau leptons produced
by neutrinos interacting just below the surface of Earth, the
neutrino event rate with a small cross section is actually larger
than the HAS rate with a large cross section;
\item the future detectors can also measure the neutrino cross section
at energies far beyond those achievable in collider experiments
thus providing important information for particle physics.
\end{enumerate}

If the proposed model is correct, the above scenario may have better 
chances to be realized.

\section{Conclusions}

UHE neutrinos became an extremely rich and interesting testing
field  for various physical phenomena, such as cosmology,
astrophysics (origin of UHE neutrinos), theory of gravitation 
(production of black holes), unified theories, testing the standard theory
and beyond, as well as the  strong interaction theory, responsible
for the hadron structure. We predict that the UHE
neutrino cross sections rise moderately with energy, in accord
with the above-mentioned ``soft'' Pomeron dynamics. 

In the present paper we concentrated on the calculation of the
asymptotic behavior of the structure functions and resulting
cross sections. We are aware, however, about the important role of
the large and intermediate $x$ behavior of the structure
functions to be fitted to the existing, relatively low-energy,
data. The cross sections calculated in this way should match with their
UHE asymptotic behavior. We intend to address this point in a
forthcoming publication.

\appendix

\section{Appendix}

In this Appendix we include formulas that are not strictly
necessary to understand the logical lines in the main text, but can
be useful in reproducing some relevant results. 

In order to determine the asymptotic behavior of $\sigma^{\nu N}$ we
needed some cumbersome integrals and the first one is
\begin{eqnarray}
&&J_-(s,Q^2)=\int_{Q^2/s}^1\frac{dx}{x}
\left[1+\left(1-\frac{Q^2}{xs}\right)^2\right]xf_q^-(x,Q^2) \nonumber \\
&= & \frac{1}{2}e^{-d_-(1)t}\left\{-3a_q+\left(3C_1(\nu)+\frac{7}{2}\right)
b_q+\frac{4Q^2}{s}[a_q-(C_1(\nu)+1)b_q] \right. \nonumber \\& + &
\frac{Q^4}{s^2}\left[-a_q+\left(C_1(\nu)+\frac{1}{2}\right)b_q\right]+4
\ln\left(\frac{s}{Q^2}\right) \left[a_q-\left(C_1(\nu)+\frac{3}{4}\right)b_q\right] 
\nonumber \\
& + & 2 \left. \ln^2\left(\frac{s}{Q^2}\right) b_q \right\}\;.
\label{a1}
\end{eqnarray}

Next, we prove that the integral
\begin{equation}
\frac{Q^2}{s}\int_{Q^2/s}^{x_M}\frac{dx}{x^2}\rho I_1(\sigma)
\label{a2}
\end{equation}
is non-leading, $O(\rho)$, with respect to the integral evaluated
in Eq.~(\ref{n15}). The identity
\begin{equation}
\frac{d}{dx}\left[\rho^n
I_n(\sigma)\right]=-\frac{1}{x}\,\rho^{n+1}I_{n+1}(\sigma),\;\;\;\;n=
0,1,2,\ldots \label{a3}
\end{equation}
can be easily obtained by using the definitions (\ref{n8}).
Integrating repeatedly by part, we get
\begin{displaymath}
\int\,\frac{dx}{x^2}\rho I_1(\sigma)=-\frac{1}{x}\rho
I_1(\sigma)-\int\,\frac{dx}{x^2} \rho^2 I_2(\sigma)=
\end{displaymath}
\begin{equation}
=\frac{1}{x}\sum_{k=1}^{\infty} (-1)^k\rho^k I_k(\sigma)\;.
\label{a4}
\end{equation}
The dominant contribution to the integral~(\ref{a2}) comes from
the lower limit of integration and, since the $z$ asymptotics of
$I_k(z)$ does not depend on $k$, we get
\begin{displaymath}
\frac{Q^2}{s}\int_{Q^2/s}^{x_M}\frac{dx}{x^2}\rho
I_1(\sigma)\approx \biggl(1+O(1/\sqrt{\ln s})\biggr) \left[\rho
I_1(\sigma)\right]_{x=Q^2/s}
\end{displaymath}
that is non-leading with respect to the integral in Eq.~(\ref{n15}). 
By using the same arguments, it is easy to show that
\begin{displaymath}
\frac{Q^4}{s^2}\int_{Q^2/s}^{x_M}\frac{dx}{x^4}\rho
I_1(\sigma)\approx \frac{1}{2}\biggl(1+O(1/\sqrt{\ln s})\biggr) \left[\rho
I_1(\sigma)\right]_{x=Q^2/s}\;.
\end{displaymath}

\vspace{0.5cm} {\bf \large Acknowledgments} L.J. thanks the
Dipartimento di Fisica Teorica dell'Universit\`a di Torino,
the Dipartimento di Fisica dell'Universit\`a della Calabria 
and the Dipartimento di Fisica dell'Universit\`a di Padova, together 
with the Istituto Nazionale di Fisica Nucleare (INFN), Sezioni di Padova e Torino 
e Gruppo collegato di Cosenza, where this work was done, 
for their warm hospitality and financial support. A.K. thanks for the same the 
Dipartimento di Fisica dell'Universit\`a di Padova and the INFN Sezione di 
Padova; he is supported by the INFN-LThPh agreement program. 
L.J. thanks A.~Kusenko for an inspiring correspondence.

\vfill \eject


\begin{thebibliography}{99}

\bibitem{YMA} Yu.M. Andreev et al., Phys. Lett. B {\bf 84} (1979) 247.

\bibitem{KR} D.W. McKay and J.P. Ralston, Phys. Lett. B {\bf 167} (1986) 103.

\bibitem{RG} R. Gandhi et al., Phys. Rev. D {\bf 58} (1998) 093009;
Astropart. Phys. {\bf 5} (1996) 81.

\bibitem{KUS} A. Kusenko and T.J. Weiler, Phys. Rev. Lett. {\bf
88} (2002) 161101; A.~Kusenko, {\tt hep-ph/0203002}.

\bibitem{KMS} J. Kwiecinski, A.D. Martin and A.M. Stasto, Phys. Rev.
D {\bf 59} (1999) 093002.

\bibitem{MR} M. Reno et al., {\tt hep-ph/0110235}.

\bibitem{MG} M. Gluck et al., Astropart. Phys. {\bf 11} (1999) 327.

\bibitem{RB} R. Basu et al., JHEP {\bf 10} (2002) 012.

\bibitem{BKK} A.V. Butkevich et al., Zeit. Phys. C {\bf 39} (1988) 241.

\bibitem{DAD} D.A. Dicus et al., Phys. Lett. B {\bf 514} (2001) 103.

\bibitem{KK03} K.~Kutak and J.~Kwiecinski, hep-ph/0303209.

\bibitem{J-M03} J.~Jalilian-Marian, hep-ph/0301238.

\bibitem{power} E.A.~Kuraev, L.N.~Lipatov and V.S.~Fadin, Sov.
Phys. JETP {\bf 45} (1977) 199; Ya.Ya.~Balitsky and L.N.~Lipatov,
Sov. J. Nucl. Phys. {\bf 28} (1978) 822 and Sov. Phys. JETP {\bf
63} (1986) 904.

\bibitem{H1} H1 Collab., C.~Adloff et al., Phys. Lett. B {\bf 520} (2001) 183.

\bibitem{BAB} L.~Babukhadia (D0 Collab.), {\tt hep-ex/0106069}.

\bibitem{FKR} G.M.~Frichter, D.W.~McKay and J.P.~Ralston, Phys. Rev. Lett. 
{\bf 74} (1995) 1508. 

\bibitem{fits} A.D. Martin et al., Eur. Phys. J. C {\bf 23} (2002)
73;  {\tt hep-ph/0211080}; S.I.~Alekhin, Phys.
Rev. D {\bf 63} (2001) 094022; V.G.~Krivokhijine and  A.V.~Kotikov, 
{\tt hep-ph/0108224}.

\bibitem{ELBA} L. Jenkovszky, F. Paccanoni and E. Predazzi, Nucl.
Phys. B (Proc. Suppl.) {\bf 25} (1992) 80.

\bibitem{DGJ} P. Desgrolard et al., Phys. Lett. B {\bf 309} (1993)
191.

\bibitem{Rivista} M. Bertini et al., Rivista Nuovo Cimento {\bf 19}
(1996) 1.

\bibitem{DJLP} P. Desgrolard et al., Phys. Lett. B {\bf 459} (1999)
265.

\bibitem{CAL} L. Csernai et al., Eur. Phys. J. C {\bf 24} (2002)
205.

\bibitem{JLP} L. Jenkovszky, A. Lengyel and F. Paccanoni, Nuovo
Cim. A {\bf 111} (1998) 551.

\bibitem{BU} A. Buras, Rev. Mod. Phys. {\bf 52} (1980) 199.

\bibitem{CCFR1} E. Oltman et al., Zeit. Phys. C {\bf 53} (1992) 51.

\bibitem{COSENZA} R.~Fiore, L.L.~Jenkovszky, E.A.~Kuraev, A.I.~Lengyel,
F.~Paccanoni and A.~Papa, Phys. Rev. D {\bf 63} (2001) 056010;
R.~Fiore, L.L.~Jenkovszky, F.~Paccanoni and A.~Papa, Phys. Rev. D {\bf 65} 
(2002) 077505.

\bibitem{FLEMING} B.T. Fleming et al. (CCFR/NuTev Coll.), Phys.
Rev. Lett. {\bf 86} (2001) 5430.

\bibitem{DGLAP} G. Altarelli and G. Parisi, Nucl. Phys. B {\bf 126}
(1977) 298; V.N. Gribov and L.N. Lipatov, Yad. Fiz. {\bf 15}
(1972) 781 [Sov. J. Nucl. Phys. {\bf 15} (1972) 438]; L.N.
Lipatov, Yad. Fiz. {\bf 20} (1974) 181 [Sov. J. Nucl. Phys. {\bf
20} (1974) 94]; Yu. L. Dokshitzer, Zh. Eksp. Teor. Fiz. {\bf 73}
(1977) 1216 [Sov. Phys. JETP {\bf 46} (1977) 641].

\bibitem{KP} A.V. Kotikov and G. Parente, Nucl. Phys. B {\bf549}
(1999) 242.

\bibitem{GR} I.S. Gradshteyn and I.M. Ryzhik, {\em Tables of Integrals, 
Series, and Products}, Academic Press. 

\bibitem{Gaz} A.Z.~Gazizov and S.I.~Yanush, Phys. Rev. D {\bf 65} (2002) 
093003, {\tt astro-ph/0112244}, {\tt astro-ph/0201528}.

\bibitem{RW} A. De Roeck and E.A. De Wolf, Phys. Lett. B {\bf 388} (1996) 843.

\bibitem{AK} A.V. Kotikov, Phys. Rev. D {\bf 49} (1994) 5746;
Phys. Atom. Nucl. {\bf 57} (1994) 133.

\bibitem{AK1} A.V.~Kotikov, Mod. Phys. Lett. A {\bf 11} (1996) 103;
Phys. Atom. Nucl. {\bf 59} (1996) 2137.

\bibitem{JKP} L.L.~Jenkovszky, A.V.~Kotikov and F.~Paccanoni,
Sov. J. Nucl. Phys. {\bf 55} (1992) 1224; 
Phys. Lett. {\bf B314} (1993) 421.

\bibitem{GLR83} L.V.~Gribov, E.M.~Levin and M.G.~Ryskin, Phys. Rep. {\bf 100} 
(1983) 1; A.H.~Mueller and J.~Qiu, Nucl. Phys. B {\bf 268} (1986) 427; 
L.~McLerran and R.~Venugopalan, Phys. Rev. D {\bf 49} (1994) 2233, D 
{\bf 49} (1994) 3352, D50 (1994) 2225; Y.V.~Kovchegov, Phys. Rev D {\bf 60} 
(1999) 034008; W.~Zhu and J.~Ruan, Nucl. Phys. B {\bf 559} (1999) 378; 
J.~Bl\"umlein et al., Phys. Lett. B {\bf 504} (2001) 235.

\end{thebibliography}
\end{document}